\documentclass[11pt,twoside]{article}
\usepackage{fancyhdr}
\usepackage[colorlinks,citecolor=blue,urlcolor=blue,linkcolor=blue,bookmarks=false]{hyperref}
\usepackage{amsfonts,epsfig,graphicx}
\usepackage{afterpage}
\usepackage{comment}
\usepackage{amsmath,amssymb,amsthm} 
\usepackage{fullpage}
\usepackage[T1]{fontenc} 
\usepackage{epsf} 
\usepackage{graphics} 
\usepackage{amsfonts,amsmath}
\usepackage[sort]{natbib} 
\usepackage{psfrag,xspace}
\usepackage{color,etoolbox}
\usepackage{ulem}

\newcommand{\cov}{\text{cov}}

\newcommand{\Pb}{\mathbb{P}}

\newcommand{\E}{\mathbb{E}}
\newcommand{\R}{\mathbb{R}}

\DeclareSymbolFont{bbold}{U}{bbold}{m}{n}
\DeclareSymbolFontAlphabet{\mathbbold}{bbold}

\theoremstyle{definition}

\theoremstyle{remark}

\begin{document}

\def\spacingset#1{\renewcommand{\baselinestretch}%
{#1}\small\normalsize} \spacingset{1}

\raggedbottom
\allowdisplaybreaks[1]


  \title{\vspace*{-.4in} {Comment on ``Statistical Modeling: The Two Cultures'' \\ by Leo Breiman}}
  \author{\\ Matteo Bonvini, Alan Mishler\thanks{MB and AM contributed equally.}, \ Edward H. Kennedy  \\ \\
    Department of Statistics \& Data Science \\
    Carnegie Mellon University \\ \\ 
    \texttt{\{ mbonvini, amishler, edward \} @stat.cmu.edu} \\
\date{}
    }

  \maketitle
  \thispagestyle{empty}

\begin{abstract}
Motivated by Breiman's rousing 2001 paper on the ``two cultures'' in statistics, we consider the role that different modeling approaches play in causal inference. We discuss the relationship between model complexity and causal (mis)interpretation, the relative merits of plug-in versus targeted estimation, issues that arise in tuning flexible estimators of causal effects, and some outstanding cultural divisions in causal inference.
\end{abstract}

\noindent

\section{Introduction}
\citet{breiman2001statistical} describes two distinct data analysis cultures that animated the statistical community at the turn of the 21st century. {One culture, popular among statisticians, starts with an explicit probabilistic model of the data generating process (DGP) and uses this model to answer questions of interest. In Breiman’s view, such probabilistic models are frequently misleading; furthermore, he argues that this approach is incapable of tackling modern challenges involving large, complex datasets, with potentially more variables than observations.}

{Breiman argues that such challenges can only be effectively solved by embracing a purely algorithmic approach. Analysts in this culture choose a method with the highest predictive accuracy, regardless of whether it corresponds to some interpretable, parsimonious model for the DGP. They then extract whatever information they can from the model post-hoc.}

In this commentary, we discuss how Breiman's points relate to problems in causal inference. Understanding the effect of X on Y is fundamentally about predicting what happens if the distribution X is artificially manipulated in some way, which results in an extra layer of complexity compared to standard prediction problems.

{We first discuss how causal misinterpretations can arise regardless of whether the data generating process is explicitly specified or not. Next, we highlight two different approaches to causal parameter estimation -- plug-in vs. targeted estimation -- that are loosely analogous to Breiman's two cultures. We then expand on the targeted approach and discuss how optimal estimation of target parameters can require undersmoothing the nuisance parameter estimates. This is a setting in which {simply} optimizing for predictive accuracy in the nuisance parameters {may} lead to suboptimal performance in the overall estimator. Finally, motivated by Breiman, we briefly note several current cultural divisions in causal inference.}

\section{Model Complexity and Causal (Mis)interpretations}


Breiman describes explicit models of the DGP as being motivated by a desire for interpretability. One of his chief complaints is that it is easy to misinterpret such models. If the model does not describe nature well, then conclusions drawn from its parameter estimates may be spurious or misleading. Implicit in Breiman's critique is the fact that misinterpretations of parameters often have a causal flavor. As Breiman alludes to, in many fields, it has historically been common to run a linear or generalized linear regression and then interpret the coefficients as representing causal effects, without consideration of whether this interpretation is justified. In Breiman's view, algorithmic modeling approaches sidestep this pitfall. By not specifying a DGP, these methods avoid presenting users with parameters which they may be tempted to misinterpret. 

We argue, however, that explicit models of the DGP are not necessarily intrinsically misleading, and that nothing in the algorithmic approach intrinsically protects against misinterpretation. Here it seems useful to distinguish several senses of the word ``model.'' In one usage, common among statisticians, ``model'' refers to a collection of probability distributions, one of which is generally assumed to represent the true data generating process. In a second usage, ``model'' refers to an algorithm that maps a dataset---the training data---to a function which itself maps from the covariate or feature space to the output space. That is, a model is a procedure used to construct a predictor from data. In a third, closely related usage, ``model'' refers to the output of this procedure in a particular instance, i.e. a fixed predictor induced by a particular training dataset.

We note that the use of a model in the second or third sense need not imply a commitment to a model in the first sense. Simple parametric predictors may be used without assuming that the regression function follows an analogous parametric form. Least squares linear regression, for example, can be understood as estimating the best linear approximation to the true regression function, regardless of the shape of that function \citep{white1982maximum, buja_models_2019_1, buja_models_2019_2}. The resulting coefficients represent predictive relationships among the variables, no matter how they are causally related. The coefficients do not need to be mapped onto any particular process in nature.

On the other side, there is a large and growing literature dedicated to rendering complex algorithmic predictors legible. This work falls under the heading of explainable, interpretable, or transparent machine learning. These methods involve constructing or extracting simple quantities that describe the behavior of a complex model. There are now many such methods available \citep{guidotti_survey_2018, arya_one_2019}. Some involve approximating the complex model by simple global or local models or decision rules; others quantify the "contribution" of each feature to predictions in order to assign importance weights to each of the features. Breiman's own feature importance method for random forests \citep{breiman_random_2001} is an example of such a method, and perhaps he would be pleased that there has been so much work in this area since.

While the proliferation of methods for interpretable machine learning speaks to the desire users have to gain insight into the behavior of complex models, the outputs of many of these methods are susceptible to exactly the same types of misinterpretations as the parameter estimates from simpler models. Indeed, many approximation methods involve simple parametric models, such as linear models, and produce precisely the same types of estimates \citep{ribeiro_why_2016}. Feature importance methods yield weights for each feature, which are superficially similar to the coefficient estimates from a linear model. Though we have not attempted to verify this, we suspect that it is common to read feature importance values as indicating relative causal importance, even when this interpretation is not licensed.

It is simple to see that the method used to model the data is orthogonal to whether or not causal inference assumptions are satisfied. Consider a causal graph $X \leftarrow Y$, where $Y$ causes $X$, and suppose we construct a model that predicts $Y$ from $X$. Whatever quantities we use to summarize the model, whether they are parameter estimates or feature importance weights or other outputs from interpretable machine learning methods, it may be tempting to read these quantities as representing the causal influence of $X$ on $Y$. This interpretation will always be invalid, because $X$ has no causal effect on $Y$.

In short, it is always possible to misread causality into a predictive model, regardless of whether the model is parametric or nonparametric, simple or complex. The solution, presumably, is to foreground the assumptions required for causal inference, rather than relying on model opacity to protect us from ourselves. After all, in many scientific settings, causal inference is the goal, whether this is expressed explicitly or not. A user who is determined to find quantities to which causal conclusions can be attached will have many ways of doing so, whatever modeling technique they use.

\section{Plug-in Versus Targeted Estimation}

Under appropriate causal identification assumptions, there are a variety of ways to proceed with estimation. The two modeling approaches that Breiman describes are roughly analogous to two distinct approaches to causal parameter estimation, which differ in whether they prioritize modeling the entire DGP or just estimating a particular parameter well. In the first approach, one starts by developing a generative model for the entire set of observed data, and then plugs in estimates from this model to answer any generic scientific question of interest. Often finite-dimensional parametric models are used, in which case the estimates takes the form of particular coefficients embedded in the model (or combinations thereof). Examples of this kind of approach include the parametric g-formula \citep{taubman2009intervening, keil2014parametric} as well as other classical approaches relying on coefficients in logistic regression or Cox models. Matching can also be viewed as a form of plug-in estimator, based on nearest-neighbor-type regression predictions \citep{abadie2006large}. There are some advantages to this kind of approach, including its simplicity and the relative interpretability of the estimators (i.e., one simply takes the estimand of interest, and plugs in estimates from the model for any unknown quantities). 

However, there are some important disadvantages to the above plug-in approach. First, simple parametric models induce the risk of substantial bias due to model misspecification. Alternatively, if one uses more flexible methods, then inference is no longer straightforward (e.g., the bootstrap may not provide valid standard errors), and in general the resulting estimator will not be efficient in a nonparametric model (i.e., there will exist different estimators with smaller mean squared error, despite not making stronger assumptions). 

These limitations motivate a second approach, where the estimation procedure is tailored to the specific scientific question and estimand of interest. For many estimands, such as average treatment effects of discrete treatments, this tailoring can be framed in terms of semi/nonparametric efficiency theory  \citep{pfanzagl1982contributions,  bickel1993efficient,van2003unified,tsiatis2006semiparametric,kennedy2016semiparametric}. Namely, for a given pathwise differentiable estimand (i.e., functional or parameter, for example the average treatment effect identified under causal assumptions such as no unmeasured confounding), one can derive the efficient influence function, which yields a benchmark for efficient estimation: no estimator can have mean squared error smaller than the variance of the efficient influence function divided by sample size, in a local asymptotic minimax sense \citep{van2002semiparametric}. The efficient influence function is also the crucial ingredient in standard recipes for constructing efficient estimators, for example based on one-step bias correction, estimating equations, or targeted maximum likelihood. 

Although in some cases the estimators themselves may not be as intuitive as simple plug-ins,  methods that are tailored to specific estimands of interest come with some important advantages. First, they often force analysts and investigators to think hard about the scientific question in formulating the estimand of interest, rather than resorting right away to modeling and outputting coefficients. More concretely, though, estimators based on influence functions (now often called doubly robust or double machine learning methods) can be optimally efficient in nonparametric models, achieving fast parametric $1/n$-type mean squared errors even in large nonparametric models, where components of the underlying data generating process can only be estimated at slower rates. In addition to the references mentioned above, we also refer to a fast-growing recent  literature for more details \citep{van2011targeted, chernozhukov2018double, robins2017minimax}.

\section{Tuning in Targeted Estimation}

One of Breiman's core arguments is that methods of data analysis should be chosen based on their out-of-sample prediction performance. In standard prediction problems, this can be accomplished, for instance, via cross-validation. In causal inference, however, the target parameters are often complex functionals that depend on several nuisance functions. Suppose these functions are estimated using an algorithm that depends on a tuning parameter $k$: how should $k$ be chosen? 

One could take the usual route and choose $k$ based on cross-validation in a way that is optimal for estimating the nuisance functions, say in terms of mean-square-error. Although this choice can still lead to nonparametric efficient estimators, it is not always optimal for estimating the functional itself, which is the ultimate target of inference. For example, consider estimating the expected density $\chi(f) = \int f^2(x) dx$, which can be estimated at root-$n$ rates even in a nonparametric model, e.g., under smoothness assumptions \citep{bickel1988estimating}. An intuitive estimator is $n^{-1}\sum_{i = 1}^n \widehat{f}(X_i)$, where, for convenience, we may think of $\widehat{f}$ as a minimax-optimal estimator of $f$ computed from a separate independent sample. A simple calculation shows that the estimation error contains the term $\int \{\widehat{f}(x) - f(x)\} f(x) dx$. In a nonparametric model, the best convergence rate for $\widehat{f}$ is always slower than root-$n$, which means that this procedure will never result in a root-$n$ consistent estimator unless some modifications are introduced.

\subsection{Influence function-based estimators}

To achieve faster convergence rates in nonparametric models, plug-in estimators of "smooth" functionals can be "corrected" by adding averages of estimates of their influence function(s)  \citep{bickel1993efficient,
van2002semiparametric, tsiatis2006semiparametric, kennedy2016semiparametric, 
chernozhukov2018double}. Suppose $\chi(\Pb) \in \R$ is smooth enough to satisfy a von Mises expansion of the form
\begin{align*}
	\chi(\widehat{\Pb}) - \chi(\Pb) = -\int \phi(o; \widehat{\Pb}) d\Pb(o) + R_2(\Pb, \widehat{\Pb})
\end{align*}
for a second order remainder term $R_2$ and a (first order) influence function $\phi(O; \Pb)$. This expansion, which is the functional analogue of a Taylor expansion, suggests a "corrected estimator" \begin{align*}
\widehat{\chi}(\widehat{\Pb}) = \chi(\widehat{\Pb}) + \frac{1}{n} \sum_{i = 1}^n \phi(O_i; \widehat{\Pb})
\end{align*}
{where $\chi(\widehat{\Pb})$ is the plug-in estimator.} We will refer to estimators based on first order influence functions simply as first order estimators, though they are also called doubly robust, targeted, or double machine learning estimators. For instance, for $\chi(\Pb) = \E\{\E(Y \mid A = 1, X) - \E(Y \mid A = 0, X)\}$, i.e. the average treatment effect under no-unmeasured-confounding, $\widehat{\chi}(\widehat{\Pb})$ takes the form of the celebrated doubly-robust estimator \citep{robins1995semiparametric, robins2000inference, van2003unified}:
\begin{align*}
\widehat{\chi}(\widehat{\Pb}) = \sum_{i=1}^n \left[ \frac{A - \widehat{\pi}(X_i)}{\widehat\pi(X_i)\{1 - \widehat{\pi}(X_i)\}}\{Y_i - \widehat\mu_A(X_i)\} + \widehat\mu_1(X_i) - \widehat\mu_0(X_i) \right]
\end{align*}
where $\mu_a(x) = \E(Y \mid A = a, X = x)$ and $\pi(x)= \Pb(A = 1 \mid X = x)$ are the outcome regression and propensity score, respectively. 

The key advantage of correcting a plug-in estimator by subtracting off an estimate of the functional's first order influence function term is that the remainder error $R_2$ is second order. For instance, in the case of the average treatment effect, we have
\begin{align*}
R_2(\Pb, \widehat{\Pb}) = -\int \{\widehat{\pi}(x) - \pi(x)\} \left\{\frac{\widehat\mu_1(x) - \mu_1(x)}{\widehat\pi(x)} - \frac{\widehat\mu_0(x) - \mu_0(x)}{1 - \widehat\pi(x)}\right\} f(x) dx
\end{align*}
where $f(x)$ is the density of $X$. If $\widehat\pi(x)$ is bounded away from zero and one, an application of Cauchy-Schwarz yields that 
\begin{align*}
| R_2(\Pb, \widehat\Pb) | \leq C \sqrt{\int \{\widehat{\pi}(x) - \pi(x)\}^2 f(x) dx} \times \max_a \sqrt{\int \{\widehat{\mu}_a(x) - \mu_a(x)\}^2 f(x) dx}
\end{align*}
for some constant $C$. Therefore, $R_2(\Pb, \widehat\Pb)$ can be $o_\Pb(n^{-1/2})$ even if the $L_2$ error in estimating the nuisance functions is $o_\Pb(n^{-1/4})$, for example. In this case, under mild conditions and provided that the nuisance functions are estimated from an independent sample, $\widehat{\chi}(\widehat{\Pb})$ would be $n^{-1/2}$-consistent and semiparametric efficient.

Achieving $n^{-1/4}$ rates for the nuisance functions, and as a result, $n^{1/2}$-consistency for the functional, generally requires that the classes in which the nuisance parameters live are sufficiently regular. For example, a $d$-dimensional regression would need to live in a $\alpha$-Holder class with $\alpha > d /2$ in order for a minimax optimal estimator to be $n^{-1/4}$-consistent \citep{tsybakov2009introduction}. Thus, if $\pi(x)$ is $\alpha$-smooth and $\mu_a(x)$ is $\beta$-smooth, $R_2(\Pb, \widehat\Pb)$ is $o_\Pb(n^{-1/2})$ if $\alpha + \beta > d$. Interestingly, there are function classes that admit estimation at rates that do not depend exponentially on $d$. For instance, functions that are cadlag and have bounded variation norm can be estimated with error $o_\Pb(n^{-1/4})$ using the Highly Adaptive Lasso \citep{van2017generally}. Similarly, certain neural network classes allow $n^{-1/4}$-consistency (up to log factors) irrespective of the dimension of the covariates \citep{gyorfi2002distribution}. 

Corrected estimators based on first-order influence functions represent a substantial improvement over plug-ins because they allow for efficient estimation even in (not overly complex) nonparametric models. Furthermore, they are straightforward to implement, as they do not require any extra tuning beyond estimating the nuisance functions. One may then wonder: can first-order corrected estimators be further improved? The answer is yes, under certain conditions, but the available estimators are not as easily implementable in practice. 

\subsection{Higher-order \& double sample-split estimators}

A natural way to improve upon first-order estimators is to correct for additional terms in the von-Mises expansion by estimating and subtracting off higher order influence function terms. However, typical functionals do not possess influence functions of order $m\geq2$, and exact corrections are not possible. Despite this, estimators based on approximate higher order influence functions (HOIFs) can improve upon first-order estimators \citep{robins2008higher, robins2009quadratic, robins2017minimax}. At a high level, approximate HOIFs of a functional $\chi(\Pb)$ can be derived as exact HOIFs of a "smoothed" functional $\widetilde\chi(\Pb)$ that possesses influence functions of all orders. One can then optimally tune the approximation bias $\chi(\Pb) - \widetilde\chi(\Pb)$ and the error in estimating $\widetilde\chi(\Pb)$. Remarkably, higher-order estimators of the average treatment effect and the expected conditional covariance are root-$n$ consistent if $\alpha + \beta > d/2$, which is a necessary condition for root-$n$ consistency in the Holder setup \citep{robins2009semiparametric}. Further, under smoothness constraints on $f$, these estimators can be minimax-optimal even outside the root-$n$ regime, i.e. when $\alpha + \beta \leq d/2$ \citep{robins2008higher, robins2017minimax}. In settings where the covariate density $f$ is non-smooth, minimax estimation is largely still an open problem, even for relatively simple causal effect functionals. Finally, HOIFs are higher order U-statistics involving kernels of projections onto suitable finite-dimensional subspaces. To the best of our knowledge, there is currently no data-driven way to choose the orders of the U-statistics and the projections, which can hinder the applicability of HOIF-based estimators in practice. 

In order to construct simpler estimators than the ones based on HOIFs, \cite{newey2018cross} have proposed using particular forms of sample splitting with the intent of estimating different components of the functional with separate, independent subsamples. For example, letting $a(X) = \E(Z \mid X)$ and $\mu(X) = \E(Y \mid X )$, a first-order estimator of the expected conditional covariance $\chi(\Pb) = \E\{\cov(Y, Z \mid X)\} = \E[Z\{Y - \mu(X)\}]$ is
\begin{align*}
\widehat\chi(\widehat\Pb) = \frac{1}{n} \sum_{i = 1}^n \{Z_i - \widehat{a}(X_i)\}\{Y_i - \widehat{\mu}(X_i)\}
\end{align*}
\cite{newey2018cross} show that if $a$ and $\mu$ are estimated with suitable series estimators constructed from separate independent samples and the average is over a third sample independent of the other two, then $\widehat\chi(\widehat\Pb)$ will be root-$n$ consistent under the minimal condition $\alpha + \beta > d/2$ when $a$ is $\alpha$-smooth and $\mu$ is $\beta$-smooth. For other functionals, such as the average treatment effect, estimators based on HOIF remain the only ones known to achieve root-$n$ consistency under the minimal condition. Recently, \cite{kennedy2020optimal} has extended the approach of \cite{newey2018cross} and obtained fast rates for estimating the conditional average treatment effect, which is a non-pathwise differentiable parameter that does not satisfy the classic von Mises expansion. 

Despite being simpler to construct than the HOIF estimators, exploiting the advantages of these estimators based on plug-ins or first-order estimators that use double sample splitting does appear to  require choosing the tuning parameters for the nuisance functions in a way that is optimal for estimating the functional, rather than just the nuisance functions. Therefore, just like HOIF-based estimators, practical implementation poses some challenges. Often, the optimal choice of the tuning parameters amounts to undersmoothing when estimating the nuisance functions. Undersmoothing has played an important role in functional estimation \citep{goldstein1992optimal, hahn1998role, hirano2003efficient, van2019efficient}. However, only the optimal order (relative to the sample size) of the tuning parameters is generally known, leaving the constants unspecified. In finite samples, choosing different constants could result in widely different estimates, and thus finding the right amount of undersmoothing in applications is difficult. Finding a data-driven method for undersmoothing remains largely an open problem. We refer to Chapter 5 in \cite{wasserman2006all} for a discussion on the practical challenges of undersmoothing, and to \cite{cui2019selective} for a recent contribution that could play an important role in tuning parameter selection for doubly robust functionals. 

Finally, the improvements over first-order estimators by either using HOIFs or carefully choosing the tuning parameters and doing double sample splitting are based on some structural assumptions. For instance, choosing the optimal order of the HOIF and the nuisance tuning parameters generally requires knowledge of the smoothness / sparsity of the nuisance functions (e.g. the order of the Holder class). Remarkably, there are exceptions to this. For instance, \cite{kennedy2020optimal} (Theorem 3) constructs an estimator of the conditional average treatment effect whose performance depends on two tuning parameters, one of which is optimized by undersmoothing as much as possible. In practice, this suggests one may then keep undersmoothing until numerical issues arise. When the optimal amount of undersmoothing is hard to obtain in finite samples, Lepski's method \citep{lepski1997optimal}, originally developed for nonparametric regression problems, selects tuning parameters in a way that can adapt to the unknown level of smoothness of the function that is estimated. A variant of it has been used for estimating the expected value of a density by \citet{gine2008simple}, and with second-order influence function estimators by \citet{liu2016adaptive}. Many open problems remain.

\section{Some Cultural Divisions in Causal Inference}

In addition to the distinction between plug-in and targeted estimation, there are a number of interesting cultural divisions that are unique to (or especially prominent in) causal inference,  which do not necessarily align with Breiman's contrast of data modeling versus algorithmic modeling. The distinctions arise in the types of targets, the formalisms used to frame assumptions and define estimands, and the techniques used for estimation. Some of these cultural divisions are: 



\begin{itemize}
    \item causal discovery versus estimation,
    \item parametric versus nonparametric methods, 
    \item graphs versus potential outcomes, 
    \item explanation versus informing policy,
    \item sensitivity analysis versus bounds (versus neither).
\end{itemize}

Although  there is much recent work that aims to bridge many of these divisions, one division remains particularly pronounced: whether the scientific question is about  discovery versus estimation. Causal discovery \citep{spirtes2000causation} usually aims at questions like: ``given some large set of variables, how can we determine which have non-zero effects on others?'' This is essentially about learning the entire underlying causal structure, e.g., in the form of a graph. In contrast, estimation problems typically focus on one or a small number of exposures, and ask only what their effect is on some outcome of particular interest, leaving unspecified much of the rest of the causal structure (for example, specific relations among confounding variables). Thus far these kinds of problems have been pursued quite separately, with philosophy and machine learning communities taking the lead on discovery, and with estimation pursued more in statistics and biostatistics and epidemiology. However, we feel there could be big benefits to more cross-disciplinary collaboration on these problems; for example, this could lead to more focus on inference and error guarantees in discovery in nonparametric models, or on using faithfulness assumptions to enlarge the kinds of questions that are typically asked in the estimation world. For more details, we refer to some excellent recent discussion by Cosma Shalizi of work by Kun Zhang at the Online Causal Inference Seminar (\href{https://youtu.be/_MVi6XzOdD0}{link}). 


\section*{References}
\vspace{-1cm}
\bibliographystyle{abbrvnat}
\bibliography{bibliography}

\end{document}